\documentclass[aps,prd,floats,amsmath,amssymb,showpacs]{revtex4}

\usepackage[utf8]{inputenc} 
\usepackage{latexsym} 
\usepackage{amsmath}  
\usepackage{amssymb}
\usepackage{hyperref}
\usepackage{color,graphicx}
\usepackage{epsfig}
\usepackage{booktabs}
\usepackage{graphicx}
\usepackage[caption=false]{subfig}

\begin{document}

%%%%%%%%%%%%%%%%%%
%%%   MACROS   %%%
%%%%%%%%%%%%%%%%%%

\newcommand{\erik}[1]{\textcolor{blue}{{\bf Erik: #1}}}
\newcommand{\miguel}[1]{\textcolor{red}{{\bf Miguel: #1}}}

%%%%%%%%%%%%%%%%%
%%%   TITLE   %%%
%%%%%%%%%%%%%%%%%

\title{Critical gravitational collapse of a non-minimally coupled scalar field}

\author{Erik Jim\'enez-V\'azquez}
\email{erjive@ciencias.unam.mx}

\author{Miguel Alcubierre}
\email{malcubi@nucleares.unam.mx}

\affiliation{Instituto de Ciencias Nucleares, Universidad Nacional
Aut\'onoma de M\'exico, A.P. 70-543, M\'exico D.F. 04510, M\'exico.}

%%%%%%%%%%%%%%%%
%%%   DATE   %%%
%%%%%%%%%%%%%%%%

\date{\today}

%%%%%%%%%%%%%%%%%%%%
%%%   ABSTRACT   %%%
%%%%%%%%%%%%%%%%%%%%

\begin{abstract}
We study the critical gravitational collapse of a massless scalar field non-minimally coupled to gravity, using a quadratic coupling function with a strength parameter $\xi$.  We concentrate on critical phenomena of type II, and determine with an accuracy of at least $10^{-12}$ the value of the critical amplitude for collapse to a black hole, as well as the values of the critical and echoing exponents. Obtaining such high accuracy in the critical amplitude requires us to do a coordinate radial transformation that effectively increases resolution near the central regions by a factor of at least $10^3$. As expected, we find that for the case of small coupling the critical behaviour is very similar to that of a minimally coupled scalar field.  On the other hand, for high coupling the dynamics become so violent that we need to introduce a special slicing condition, known as the shock-avoiding slicing condition, in order to avoid gauge pathologies that would otherwise cause our simulations to fail. With this new gauge condition we are able to perform high accuracy simulations even in the strong coupling regime, where we find that the critical and echoing exponents become significantly smaller, and that the echoing behavior is richer and can not be modelled by a single harmonic.
\end{abstract}

%%%%%%%%%%%%%%%%
%%%   PACS   %%%
%%%%%%%%%%%%%%%%


%%%%%%%%%%%%%%%%%%%%%%
%%%   MAKE TITLE   %%%
%%%%%%%%%%%%%%%%%%%%%%

\maketitle

%%%%%%%%%%%%%%%%%%%%%%%%
%%%   INTRODUCTION   %%%
%%%%%%%%%%%%%%%%%%%%%%%%

\section{Introduction}

Studying the gravitational collapse of a real massless scalar field, Choptuik discovered a phenomena where at the threshold of black hole formation Einstein's field equations have a universal solution, independent of the initial data \cite{PhysRevLett.70.9}. Further studies have shown two different types of critical collapse: type I where the black holes have a finite minimum mass, and type II where taking a uni-parametric family of initial data parametrized by $p$, the black hole mass follows a scaling relation of the form:
\begin{equation}\label{eq:mass_scaling}
M\propto(p-p^*)^\gamma \,,
\end{equation}
where a black hole forms for $p>p^*$. The critical exponent $\gamma$ is universal with respect to different families of initial data, depending only on the type of matter. The exact value $p=p^*$ corresponds to a large curvature solution called the ``critical solution'', which divides the parameter space into two regimes: if $p<p^*$ the original matter content disperses and no black hole is formed, and if $p>p^*$ a black hole is always formed.
	
The critical solutions also have the property of self-similarity. This symmetry can be either continuous self-similarity (CSS), or discrete self-similarity (DSS). In the discrete case, there exists an echoing exponent $\Delta$ such that in logarithmic time:
\begin{equation}\label{eq:log_time}
T=-\ln(\tau^*-\tau)\,,
\end{equation}
the critical solution is periodic in $T$ with period $\Delta$. In equation~\eqref{eq:log_time} $\tau$ is so measure of time and $\tau^*$ is the so-called "accumulation time", i.e. a point where for every small neighbourhood there exist infinitely many echoes as we approach from the left $\tau\rightarrow(\tau^*)^-$. Usually $\tau$ is taken as the proper time of an observer located at the origin. For the case of a real massless scalar field the critical solution has a DSS, and the values of the characteristic exponents have been found to be $\gamma=0.374$ and $\Delta=3.445$ \cite{PhysRevLett.70.9,Rinne:2020asi,PhysRevD.49.890,PhysRevD.51.5558,Hamad__1996,PhysRevD.92.084037}.
The DSS found for the case of real massless scalar field is the only possibility when the scalar field is coupled minimally to gravity, but Liebling and Choptuik have shown that in the Brans-Dicke theory, the critical collapse of a scalar field can show both CSS and DSS depending on the value of the coupling parameter~\cite{PhysRevD.56.3466}.
	
In the spirit of investigating the critical collapse of a real massless scalar field in alternative theories of gravity, we will explore this phenomenon for a generalization of the Brans-Dicke theory, namely the scalar-tensor theories of gravity.  Among the many possible modifications to standard general relativity, scalar-tensor theories (STT) are those for which in the so-called Jordan frame representation a scalar field is non-minimally coupled to the Ricci scalar in such a way that it can be understood as a local variation of Newton's constant. Due to these modifications, in STT some phenomena arise that are not present in the standard general relativity description, such as an extra longitudinal component of gravitational waves~\cite{PhysRevD.50.7304,PhysRevD.55.2024}, and also a phenomenon known as ``spontaneous scalarization'', where self-gravitating solutions with no initial scalar field can spontaneously acquire a non-trivial scalar field that drives the solution into an energetically more favoured state.  This spontaneous scalarization has been studied for both neutron stars~\cite{PhysRevD.57.4789,PhysRevD.58.124003} and boson stars~\cite{PhysRevD.86.104044,DEGOLLADO2020135666}. In a cosmological scenario, STT have been proposed as dark energy models in substitution of the cosmological constant \cite{PhysRevLett.85.2236, PhysRevLett.86.196, PhysRevD.71.083512}.

This paper is the organized as follows. In Section~\ref{basics} we present a brief review of scalar-tensor theories of gravity and relevant $3+1$ equations. Section~\ref{numerical_setup} presents our numerical methods, initial data and diagnostics required to find the critical amplitude. We present the results of our numerical simulations in section~\ref{numerical_results}, and we conclude in Section~\ref{conclusion}.

%%%%%%%%%%%%%%%%%%%%%
%%%   EQUATIONS   %%%
%%%%%%%%%%%%%%%%%%%%%

\section{Basic equations}
\label{basics}

In this section we summarize the relevant equations that describe the STT of gravity (or more details see for example~\cite{Salgado:2005hx,PhysRevD.86.104044}). The action for scalar-tensor theories with a single scalar field in the Jordan frame is given by (in units such that $c=1$):
\begin{equation}\label{eq:STT-action}
S=\int \left[\frac{F(\phi)}{16 \pi G_0} \: R - \left(\frac{1}{2} g^{\mu \nu}\left(\partial_{\mu} \phi\right)\left(\partial_{\nu} \phi\right) + V(\phi)\right)\right] \, \sqrt{-g}\, d x^{4}
+ S_{\text{mat}}\left(\Psi, g_{\mu \nu}\right) \,,
\end{equation}
with $\phi$ the non-minimally coupled scalar field, $V(\phi)$ a self-interaction potential, $R$ the Ricci scalar of the spacetime,,and where $S(\Psi,g_{\mu \nu})$ represents the action of all the matter fields other than $\phi$. Finally, the function $F(\phi)$ is the non-minimally coupling function, which we have taken a to be a quadratic function of the form:
\begin{equation}\label{eq:nonmin_coupling}
F(\phi)=1+8\pi\xi G_0\phi^2\,,
\end{equation}
with $\xi$ a positive constant. This form of $F(\phi)$ has been used as a toy model for the study of scalarization in both neutron and boson stars~\cite{PhysRevD.86.104044,DEGOLLADO2020135666}. In the following we will always consider $V(\phi)=0$, corresponding to a real massless non-interacting scalar field. The constant $G_0$ is the usual gravitational constant, but notice that
in~\eqref{eq:STT-action} we can identify the ``effective'' gravitational constant as $G_{\text{eff}}=G_0/F(\phi)$. 

Varying the action with respect to the spacetime metric and the scalar field one obtains the following field equations:
\begin{eqnarray}
R_{a b}-\frac{1}{2} g_{a b} R&=& 8\pi G_{0} T_{a b} \label{eq:Einstein-STT}\,, \\
\square \phi+\frac{1}{2} f^{\prime} R&=&0 \label{eq:Klein-Gordon_STT}\,,
\end{eqnarray}
where here $f^\prime$ indicates the derivative of $f$ with respect to the scalar field $\phi$, and $\square=g^{a b}\nabla_a \nabla_b$ is the d'Alambertian operator. The effective stress-energy tensor $T_{ab}$ has three separate contributions given by:
\begin{eqnarray}
T_{a b} &:=& \frac{G_{\text {eff }}}{G_{0}}\left(T_{a b}^{f}+T_{a b}^{\phi}+T_{a b}^{\text {matt }}\right)\,, \label{eq:Tab_STT}\\
T_{a b}^{f} &:=& \nabla_{a}\left(f^{\prime} \nabla_{b} \phi\right)-g_{a b} \nabla_{c}\left(f^{\prime} \nabla^{c} \phi\right)\,,\\
T_{a b}^{\phi} &:=& \left(\nabla_{a} \phi\right)\left(\nabla_{b} \phi\right)-\frac{g_{ab}}{2}(\nabla \phi)^{2}\,, \\
G_{\text {eff }} &:=& \frac{1}{8 \pi f} \,, \quad f:=\frac{F}{8 \pi G_{0}} \,, \label{eq:G_eff}
\end{eqnarray}
where $T^{\text{mat}}_{ab}$ is the stress-energy tensor of the matter fields other than $\phi$. Taking the trace of equation~\eqref{eq:Einstein-STT}, one can rewrite the source term in~\eqref{eq:Klein-Gordon_STT} as:
\begin{equation}\label{eq:Klein-Gordon_STT_2}
\square \phi=\frac{f^{\prime} T_{\text {matt }}-f^{\prime}\left(1+3 f^{\prime \prime}\right)(\nabla \phi)^{2}}{2 f\left(1+3 f^{\prime 2} / 2 f\right)} \,,
\end{equation}
with $T_{\text{mat}}$ the trace of $T^{ab}_{\text{mat}}$. In this context, the Bianchi identities directly imply that the full stress-energy tensor is conserved:
\begin{equation}
\nabla_{c} T^{c a}=0 \,.
\end{equation}
Furthermore, the field equations also lead to the conservation of the stress-energy tensor of matter on its own:
\begin{equation}
\nabla_{c} T_{\text {matt }}^{c a}=0 \,,
\end{equation}
which implies that test particles follow the geodesics associated with the metric $g_{ab}$.

%%%%%%%%%%%%%%%%%%%%%%%%%%%%%
%%%   3+1 DECOMPOSITION   %%%
%%%%%%%%%%%%%%%%%%%%%%%%%%%%%

\subsection{3+1 decomposition}

In order to study the evolution in time of our system, we first need to recast the field equations~\eqref{eq:Einstein-STT} and~\eqref{eq:Klein-Gordon_STT} as a Cauchy problem. We do this by following the 3+1 formalism~\cite{Alcubierre:1138167}, and considering a globally hyperbolic spacetime which is foliated by a family of spacelike hypersurfaces $\Sigma_t$ parametrized by a global time function $t$. With these considerations, we rewrite the spacetime metric $g_{ab}$ in the form:
\begin{equation}
d s^{2}=-\left(\alpha^{2}-\beta^{i} \beta_{i}\right) d t^{2}+2 \beta_{i} d x^{i} d t+\gamma_{i j} d x^{i} d x^{j} \,,
\end{equation}
with $\alpha$ the lapse function, $\beta^i$ the shift vector, and $\gamma_{ij}$ the 3-metric induced on the spatial hypersurfaces. We introduce the unit normal timelike vector $n^a$ to the spacelike hypersurfaces $\Sigma_t$, and perform the $3+1$ decomposition via the projection operator $P^a_{\ b}:=\delta^a_{\ b}+n^a n_b$. The extrinsic curvature $K_{ab}$ of the spatial hypersurfaces $\Sigma_t$ is then given by:
\begin{equation}
K_{a b}:=-P_{\ a}^{c} \nabla_{c} n_{b}=-\left(\nabla_{a} n_{b}+n_{a} n^{c} \nabla_{c} n_{b}\right) \,.
\end{equation}
For the scalar field it is convenient to introduce the following auxiliary quantities:
\begin{eqnarray}
Q_{i} &:=& D_{i} \phi=P_{i}^{k} \nabla_{k} \phi \,,\label{eq:Q}\\
\Pi &:=& n^{a} \nabla_{a} \phi= \frac{1}{\alpha} \frac{d \phi}{d t} \,,
\end{eqnarray}
where $D_i$ is the covariant derivative compatible with the 3-metric $\gamma_{ij}$, and where we have defied the operator \mbox{$d/dt:=\partial_{t}-\mathcal{L}_{\beta}$}, with $\mathcal{L}_{\beta}$ the Lie derivative along the shift vector. The evolution equations for $Q_i$ and $\Pi$ then become:
\begin{eqnarray}
\frac{d Q_{i}}{d t} &=& D_{i}(\alpha\Pi) \,,
\label{eq:Qdot} \\
\frac{d \Pi}{d t} &=& \alpha \left[ \Pi K+Q^{l} D_{l}(\ln \alpha) + D_{l} Q^{l} \right] \nonumber \\
&-& \frac{\alpha}{2 f \left( 1 + 3 {f^\prime}^2 /2 f \right)} \left[ 2 f V^{\prime}-4 f^{\prime} V - f^{\prime}\left(1+3 f^{\prime \prime}\right)\left(Q^{2}-\Pi^{2}\right)+f^{\prime} T_{\mathrm{matt}}\right] \,.
\label{eq:dPi_dt}
\end{eqnarray}

From the orthogonal decomposition of the stress-energy tensor:
\begin{equation}
T^{a b}=S^{a b}+J^{a} n^{b}+n^{a} J^{b}+\rho n^{a} n^{b} \,,
\end{equation}
we obtain the energy density $\rho:=n^{a} n^{b} T_{a b}$, the momentum density $J^a:= -P_{\ a}^{b} n^{c} T_{b c}$, and the stress tensor \mbox{$S^{ab}:=P_{a}^{c} P_{b}^{d} T_{c d}$}. From \eqref{eq:Tab_STT} we see that each component of the stress-energy tensor has three separate contributions:
\begin{eqnarray}
\rho &=&\frac{G_{\text {eff }}}{G_{0}}\left(\rho^{f}+\rho^{\phi}+\rho^{\text {matt }}\right) \,,\label{eq:energy-density} \\
J_{i} &=&\frac{G_{\text {eff }}}{G_{0}}\left(J_{i}^{f}+J_{i}^{\phi}+J_{i}^{\text {matt }}\right), \label{eq:momentum-density} \\
S_{i j} &=&\frac{G_{\text {eff }}}{G_{0}}\left(S_{i j}^{f}+S_{i j}^{\phi}+S_{i j}^{\text {matt }}\right) . \label{eq:stress-tensor}
\end{eqnarray}
Using now equations \eqref{eq:G_eff} and \eqref{eq:Klein-Gordon_STT_2}, the explicit expressions for the matter quantities \eqref{eq:energy-density}-\eqref{eq:stress-tensor} become:
\begin{eqnarray}
\label{eq:rho_STT}
\rho &=& \frac{1}{8 \pi G_{0} f} \left[ f^{\prime}\left(D_{k} Q^{k}
+ K \Pi\right) + \frac{\Pi^{2}}{2}
+ \frac{Q^2}{2} \left( 1 + 2 f^{\prime \prime} \right)
+ V(\phi) + \rho_{\text {matt }} \right] \,, \\
\label{eq:J_STT} 
J_{i} &=& \frac{1}{8 \pi G_{0} f} \left[ - f^{\prime} \left( K_{i}^{k} Q_{k} + D_{i} \Pi \right)
- \Pi Q_{i} \left( 1 + f^{\prime \prime} \right) + J_{i}^{\mathrm{matt}} \right] \,, \\
S_{i j} &=& \frac{1}{8 \pi G_{0} f} \left\{ \rule{0mm}{5mm}
Q_{i} Q_{j} \left( 1 + f^{\prime \prime} \right)
+ f^{\prime} \left( D_{i} Q_{j} + \Pi K_{i j} \right) \right. \nonumber \\
&-& \frac{\gamma_{i j}}{\left(1 + 3 {f^\prime}^2/2f\right)}
\left[ \frac{1}{2} \left( Q^{2 }- \Pi^{2} \right)
\left( 1 + \frac{f^{\prime 2}}{2f} + 2 f^{\prime \prime} \right) \right. \nonumber \\
&+& \left. \left. V \left( 1 - {f^\prime}^2/2f \right)
+ f^{\prime} V^{\prime} + \frac{f^{\prime 2}}{2f}
\left( S_{\text {matt}} - \rho_{\text {matt}} \right) \right]
+ S_{i j}^{\text {matt }}\right\} \,, \label{eq:S_STT}
\end{eqnarray}
with $Q^2:=Q^lQ_l$, and where $K:=\gamma^{ij}K_{ij}$ is the trace of the extrinsic curvature tensor.

The $3+1$ evolution equations obtained from the field equation~\eqref{eq:Einstein-STT} are the standard ADM equations, given by:
\begin{eqnarray}
\frac{d \gamma_{i j}}{d t} &= & -2 \alpha K_{i j} \,, \label{eq:dgij_dt}\\
\frac{d K_{i j}}{d t} &=& - D_{i} D_{j} \alpha
+ \alpha \left[ R_{i j} + K K_{i j} - 2 K_{i l} K_{j}^{l} \right] \nonumber \\
&+& 4 \pi G_{0} \alpha \left[ \gamma_{ij} (S - \rho) - 2 S_{i j} \right] \,,
\label{eq:dkij_dt}
\end{eqnarray}
where $R_{ij}$ is the 3-Ricci tensor associated with the spatial metric $\gamma_{ij}$, and $S:=\gamma^{ij}S_{ij}$. The hamiltonian and momentum constraints take the form:
\begin{eqnarray}
H &:=&\frac{1}{2}\left( R + K^{2} - K_{i j} K^{i j} \right) - 8 \pi G_{0} \rho = 0 \,,
\label{eq:ham_constraint}\\
M^{i} &:=& D_{l} \left( K^{il} - \gamma^{il} K \right) - 8 \pi G_{0} J^{i} = 0 \,. \label{eq:mom_constraint}
\end{eqnarray}

Finally, since above we have defined the auxiliary variable $Q_i$, formally we also need to add its definition~\eqref{eq:Q} and an integrability condition as new constraints:
\begin{eqnarray}
Q_{i}-D_{i} \phi &=0 \,,\\
D_{[i} Q_{j]} &=0 \,.
\end{eqnarray}

%%%%%%%%%%%%%%%%%%%%%%%%%%%%
%%%   GAUGE CONDITIONS   %%%
%%%%%%%%%%%%%%%%%%%%%%%%%%%%

\subsection{Gauge conditions}

Additionally to the evolution equations for the gravitational and scalar fieldsq, in order to obtain a closed evolution system we also have to impose gauge conditions for the lapse $\alpha$ and shift vector $\beta^i$. Following~\cite{PhysRevD.86.104044}, we will use a modified Bona-Masso slicing condition for the lapse given by:
\begin{equation}
\frac{d \alpha}{dt} = - \alpha^{2} F_{B M}(\alpha)
\left[ K -  \frac{\Theta}{f_{B M}(\alpha)} \: \frac{f^{\prime}}{f} \: \Pi \right] \,,
\end{equation}
with $F_{BM}(\alpha)$ a positive but otherwise arbitrary function of $\alpha$, and $\Theta$ an arbitrary parameter. The specific values $F_{BM}(\alpha)=\Theta=1$ correspond to the so-called ``pseudo-harmonic'' foliation, and have been used in the hyperbolicity analysis in~\cite{Salgado:2005hx,PhysRevD.77.104010}. With $\Theta=0$ one recovers the usual Bona-Masso slicing condition~\cite{Bona94b}, but as shown in~\cite{PhysRevD.77.104010} in our case this choice does not lead to a strongly hyperbolic formulation. For this reason, in what follows we will always take $\Theta=1$.

In relation to the choice of the Bona-Masso gauge function $F_{BM}(\alpha)$, one can take: 
\begin{equation}
F_{BM}(\alpha)=2/\alpha \; ,
\label{eq:1+log}
\end{equation}
which corresponds to the standard $1+\log$ slicing. However, in references~\cite{PhysRevD.55.5981,Alcubierre:2002iq} one of the authors (MA) explored the alternative choice:
\begin{equation}
F_{BM}(\alpha) = 1 + \kappa/\alpha^2 \; ,
\label{eq:shockavoid}
\end{equation}
with $\kappa$ a positive but otherwise arbitrary constant. This choice for $F_{BM}(\alpha)$ is made in order to avoid a particular type of gauge pathologies that lead to singular solutions. These pathologies resemble the shock waves of hydrodynamics, and for this reason are known as ``gauge shocks''. As we will show below, for some values of the non-minimal coupling constant the evolution using the 1+log slicing develops a gauge pathology that causes the numerical code to fail. We have found that these gauge pathologies can be eliminated using the gauge function~\eqref{eq:shockavoid}. In our simulations below we use this shock-avoiding gauge condition with $\kappa=1$, so that when $\alpha \rightarrow 1$ in the asymptotic region we have $F_{BM}\rightarrow2$, and our gauge condition mimics the standard $1+\log$ slicing.

Concerning the choice of the shift vector $\beta^i$ we simply set it to zero since we are mainly interested on sub-critical evolutions.  For the super-critical case when a black hole forms, a non-zero shift would be preferable in order to avoid the well-known slice stretching effects.

%%%%%%%%%%%%%%%%%%%%%%%%%%%%%%
%%%   SPHERICAL SYMMETRY   %%%
%%%%%%%%%%%%%%%%%%%%%%%%%%%%%%

\subsection{Evolution in spherical symmetry}

It is well known that the standard ADM formalism results in a weakly hyperbolic formulation of general relativity~\cite{Alcubierre:1138167}. Because of this, for our simulations we will use the BSSN formulation \cite{PhysRevD.59.024007,PhysRevD.52.5428}. As we are only considering the case of spherical symmetry, we use the generalized version of BSSN formulation which is adapted to curvilinear coordinates~\cite{Alcubierre:2011pkc,PhysRevD.86.104044}~\cite{PhysRevD.79.104029}. Under this assumptions, the conformal 3-metric decomposition takes the form:
\begin{equation}\label{eq:sph_metric}
dl^2=\psi(t,r)^4\left[A(t,r)dr^2+B(t,r)r^2d\Omega^2\right] \,,
\end{equation}
where $d\Omega^2=d\theta^2+\sin\theta^2d\varphi^2$ is the solid angle element. The evolution is performed using the spherically-symmetric BSSN version of equations~\eqref{eq:dgij_dt} and~\eqref{eq:dkij_dt},  where the energy density, momentum density and stress tensor are given by \eqref{eq:rho_STT}, \eqref{eq:J_STT} and \eqref{eq:S_STT}. Additionally, as we are expecting a spacetime with large gradients of curvature, we will use the puncture method \cite{PhysRevLett.96.111101} by evolving $\chi=\psi^{-4}$. The Klein-Gordon equation~\eqref{eq:Klein-Gordon_STT_2} is rewritten as a first order PDE system using \eqref{eq:dPi_dt}. Specifically, we evolve the metric quantities $A$, $B$, $\chi$, the trace of the trace of the extrinsic curvature $K$, the traceless part of the conformal extrinsic curvature, and
the radial component of the conformal connection functions.

%%%%%%%%%%%%%%%%%%%%%%%%%%%
%%%   NUMERICAL SETUP   %%%
%%%%%%%%%%%%%%%%%%%%%%%%%%%

\section{Numerical setup}
\label{numerical_setup}

Numerical simulations are performed using the OllinSphere code presented in \cite{PhysRevD.86.104044,Alcubierre:2014joa,PhysRevD.81.124018}. OllinSphere uses a finite difference method to discretize the Einstein's field equations using an equally spaced mesh in $r$. Following~\cite{PhysRevD.92.084037,Rinne:2020asi}, we propose a change of coordinates from the original radial coordinate $r$ to a new radial coordinate $\tilde{r}$ which is defined via the differential relation:
\begin{equation}\label{eq:rtrans}
	\frac{dr}{d\tilde{r}}=\frac{1}{1+e^{(\beta \tilde{r}^2+\delta)}} \,,
\end{equation}
where $\beta$ and $\delta$ are arbitrary constants such that $\beta<0$ and $\delta>0$. With this transformation, as $\tilde{r}$ approaches infinity we have $dr/d\tilde{r} \rightarrow 1$. On the other hand, as $\tilde{r}\rightarrow0$, the relation \eqref{eq:rtrans} approaches:
	\begin{equation}\label{rtrans_taylor}
	\frac{dr}{d\tilde{r}}=\frac{1}{1+e^\delta}-\frac{\beta e^\delta \tilde{r}^2}{(1+e^\delta)^2}+{\cal O}(\tilde{r}^4) \,,
\end{equation}
showing that the parameter $\delta$ adjusts the resolution near the origin $\tilde{r}=0$, while $\beta$ measures how fast $dr/d\tilde{r}$ approaches $1$ far away.  The typical values we use for our simulations are $\delta=5$, $\beta=-1$. With these choices, a uniform grid on $\tilde{r}$ becomes non-uniform in $r$, gaining a factor of about $10^3$ times more resolution close to the origin. One final comment related to equation \eqref{eq:rtrans}. As this expression is not analytically integrable, the differential relation must be solved numerically. In order to reduce numerical error up to machine precision we integrate this equation with a Chebyshev quadrature starting from the origin, using a fifth order Chebyshev polynomial between each grid point.

Using the change of coordinates given by \eqref{eq:rtrans} does not require any change in the internal structure of the OllinSphere code, as it already uses the most general form of the line element in spherical symmetry for the conformal metric:
\begin{equation}\label{eq:ds}
ds_3^2 = A(t,r) dr^2 + B(t,r)r^2 d\Omega^2\,,
\end{equation}
with $A$ and $B$ positive metric functions, and $d\Omega^2=d\theta^2+\sin\theta^2d\varphi^2$ the solid angle element. Once we have some initial data (see next Section), changing the radial coordinate from $r$ to $\tilde{r}$ modifies the explicit values of the metric coefficients $A$ and $B$,  but the new metric has exactly the same form as above with new metric coefficients given by:
\begin{equation}
\tilde{A} := A \left(\frac{dr}{d\tilde{r}}\right)^2 \: , \qquad
\tilde{B} := B \left( \frac{r}{\tilde{r}} \right)^2 \: .
\end{equation}

%%%%%%%%%%%%%%%%%%%%%%%%
%%%   INITIAL DATA   %%%
%%%%%%%%%%%%%%%%%%%%%%%%

\subsection{Initial data}

As mentioned before, the matter content in our numerical simulation consists of a massless scalar field coupled non-minimally to gravity. For the scalar field we consider the following initial data profiles:
\begin{eqnarray}
\phi_I(0,r) &=& \phi_0 \: e^{(-r^2/\sigma^2)} \,, \label{eq:gaussian} \\
\phi_{II}(0,r) &=& \phi_0 \: r^2 e^{(-r^2/\sigma^2)} \,,\label{eq:r2gaussian}\\
\phi_{III}(0,r) &=& \phi_0 \: \coth \left( s_0/\sigma \right)
\left[ \tanh \left( \frac{r+s_0}{\sigma} \right)
- \tanh \left( \frac{r-s_0}{\sigma} \right) \right] \,, \label{eq:tophat}
\end{eqnarray}
where $r,\sigma,s_0$ are free parameters. In our numerical simulations we fix $\sigma=1$, $s_0=0.5$, and use the amplitude $\phi_0$ as the tuning parameter for the initial pulse.

The conformal metric is initialized to the flat metric in spherical symmetry, that is $A=B=1$. However, once we change to the rescaled radial coordinate $\tilde{r}$ this implies that:
\begin{eqnarray}
\tilde{A} = \left( dr/d\tilde{r} \right)^2 \: , \qquad \tilde{B} = ( r/\tilde{r} )^2 \: .
\end{eqnarray}

We also assume time-symmetric initial data, which implies that the momentum constraint \eqref{eq:mom_constraint} is trivially satisfied.  This leaves the hamiltonian constraint \eqref{eq:ham_constraint} as the only equation to solve for the initial conformal factor $\psi(\tilde{r})$. Boundary conditions for $\psi$ are obtained from the asymptotic flatness condition:
\begin{equation}
\psi(\tilde{r})|_{\tilde{r}\rightarrow\infty}=1 \, .
\end{equation}
In practice, however, we use a boundary condition at a finite radius of the form:
\begin{equation}
\partial_{\tilde{r}}\psi = \frac{1-\psi}{\tilde{r}} \: ,
\end{equation}
which is a Robin type boundary condition and reflects the fact that as $\tilde{r}\rightarrow\infty$ we have $\psi\rightarrow1+{\cal{O}}(r^{-1})$.   
At the origin, we demand that $\psi$ must be an even function in $\tilde{r}$ for regularity, that is:
\begin{equation}
	\partial_{\tilde{r}}\psi(\tilde{{r}})|_{\tilde{r}=0}=0 \,.
\end{equation}
Additionally, the initial gauge is completely specified by choosing a pre-collapsed lapse of the form $\alpha=\psi^{-2}$, as well as zero shift vector $\beta^i=0$.

%%%%%%%%%%%%%%%%%%%%%%%
%%%   DIAGNOSTICS   %%%
%%%%%%%%%%%%%%%%%%%%%%%

\subsection{Diagnostics}

The final state of the evolution is determined by the strength of the initial data. For a critical value of the amplitude $\phi_0^*$ one finds that weak initial data with $\phi_0<\phi_0^*$ completely disperse leaving behind Minkowski spacetime, while for stronger initial data with $\phi_0>\phi_0^*$ the scalar field collapses to form black hole. The critical value $\phi_0^*$ is found using a bisection method, bracketing the interval between an amplitude for which the scalar field is dispersed $\phi_d$, and an amplitude $\phi_c$ for which a black hole forms. In a similar way to \cite{PhysRevD.92.084037}, the dimensionless quantity:
\begin{equation}
\delta \phi =\frac{\phi_c-\phi_d}{\phi_d} \,,
\end{equation}
indicates the precision with which we have found the critical amplitude. In order to obtain the critical exponents we need an accuracy equal or better than $\delta \phi\sim 10^{-6}$. Increasing precision leads to longer evolutions near the critical solution, resulting in less uncertainty in the estimation of the critical exponent $\gamma$ and echoing parameter $\Delta$.

The final state of the evolution is analyzed looking at the the behavior of the lapse at the origin. If the initial data is dispersed, the lapse will return to one as the spacetime approaches Minkowski. On the other hand, if a black hole forms our gauge condition causes the lapse to collapse to zero at the center. In order to better study the collapsing configurations, we also search for an apparent horizon at every time step. This is done by calculating the expansion of outgoing null geodesics and looking for a place where it becomes zero. In spherical symmetry this expansion takes the form:
\begin{equation}
\Theta=\frac{1}{\psi^{2} \sqrt{A}}\left(\frac{2}{r} + \frac{\partial_{r} B}{B}+4 \frac{\partial \psi}{\psi}\right) - 2 K_{\theta}^{\theta} = 0 \,,
\end{equation}
where $K_{\theta}^{\theta}$ is the angular component of the extrinsic curvature with mixed indices. Since our study only focuses on subcritical evolutions, we do not need to determine very accurately the final mass of the formed black hole.

In the subcritical regime, since $\xi=0$ corresponds to the minimally coupled case, we expect that the maximum value of the 4D-Ricci scalar evaluated at the origin will follow a scaling law corresponding to a type II critical collapse:
\begin{equation}
R_{max}\simeq|\phi_0^* - \phi_0|^{-2\gamma} \, ,
\end{equation}
where the factor $-2$ in the scaling exponent is there because the 4D-Ricci scalar has units of length to the minus two. Additionally to this behavior, Hod and Piran \cite{PhysRevD.55.R440} noticed that for a scalar field coupled minimally to gravity, the self-discrete nature of the critical phenomena adds a periodic modulation to the scaling law. The 4D-Ricci scalar is then expected to behave as:
\begin{equation}
\ln R_{max} = c - 2\gamma\ln|\phi_0^*-\phi_0| + f(\ln|\phi_0^*-\phi_0|) \,,
\end{equation}
with $c$ a constant that depends on the initial data family, and where $f(x)$ is a periodic function with a frequency given by:
\begin{eqnarray}
\omega=\frac{\Delta}{2\gamma} \,,
\end{eqnarray}
with $\Delta$ the so-called echoing exponent. Usually, to first order one can approximate $f(x)$ by a simple trigonometric function, for example: 
\begin{equation}
f(x) = a_0 \sin(\omega x+\varphi)\,.
\end{equation}
The 4D-Ricci then behaves as:
\begin{equation}\label{eq:Ricci_scaling}
\ln R_{max} = c - 2\gamma\ln|\phi_0^*-\phi_0|+a_0\sin(\omega\ln|\phi_0^*-\phi_0|+\varphi_0) \,,
\end{equation}
where the constants $c,a_0,\varphi_0$ are family dependent. Fitting the function \eqref{eq:Ricci_scaling} provides us with a first method to compute the echoing exponent $\Delta$. But there is a second method one can use to find $\Delta$ due to Baumgarte \cite{PhysRevD.98.084012}. One can consider the times for two pairs of consecutive zero crossings of the scalar field $\phi$ evaluated at the origin, $(\tau_n,\tau_{n+1})$ and $(\tau_m,\tau_{m+1})$. Substituting these values in the logarithmic time \eqref{eq:log_time} we will then have the corresponding pairs $(T_n,T_{n+1})$, $(T_m,T_{m+1})$. Assuming now that each pair differs in half the period $\Delta/2$, one can solve for the accumulation time $\tau^*$ obtaining:
\begin{equation}\label{eq:acc_t}
\tau^{*}=\frac{\tau_{n} \tau_{m+1}-\tau_{n+1} \tau_{m}}{\tau_{n}-\tau_{n+1}-\tau_{m}+\tau_{m+1}} \, .
\end{equation}
This estimation for the accumulation time also provides us with an estimate of the echoing period $\Delta$ given by:
\begin{equation}\label{eq:delta_echo}
\Delta = 2 \ln \left(\frac{\tau^{*}-\tau_{n}}{\tau^{*}-\tau_{n+1}}\right) \,.
\end{equation}

%%%%%%%%%%%%%%%%%%%%%%%%%%%%%
%%%   NUMERICAL RESULTS   %%%
%%%%%%%%%%%%%%%%%%%%%%%%%%%%%

\section{Numerical results}
\label{numerical_results}

All our simulations have been performed using a method of lines with a fourth order Runge-Kutta integration in time, and fourth order centered finite differences in space. Values for the non-minimal coupling constant $\xi$ in~\eqref{eq:nonmin_coupling} were chosen in a logarithmic scale, taking as specific values $\xi=10^{-3}, 10^{-2}, 10^{-1}, 1, 10$. For our simulations we use a grid spacing of $\Delta r = 0.005$, with $N_r=2800$ points in radial direction, and the parameters used by the radial coordinate transformation are $\delta=5$ and $\beta=-1$.  Additionally, we have used an adaptive time step in order to always satisfy the Courant-Friedrich-Levy condition required for numerical stability \cite{Alcubierre:1138167}.

Critical phenomena can be strongly affected by numerical error due to either the boundary conditions or the finite difference method. The first source of error can be highly reduced by using constraint preserving boundary conditions. These have been implemented using the algorithm described in \cite{PhysRevD.86.104044, Alcubierre:2014joa}, reducing by a factor of about $10^3$ the error introduced by the artificial boundary in comparison with the standard Sommerfeld (radiative) boundary conditions. In relation to the error introduced by the finite difference method, we also use sixth-order Kreiss-Oliger dissipation in order to be compatible with the fourth order discretization. This artificial dissipation dampens high frequency modes which would otherwise spoil the evolution near the black hole formation threshold.

For each value of the coupling constant $\xi$ we test the three different families of initial data given by equations \eqref{eq:gaussian}, \eqref{eq:r2gaussian} and \eqref{eq:tophat}. Reported values for the critical exponents $(\gamma,\Delta)$ are the averages of the critical exponents obtained for each family, and the uncertainty is taken as the highest deviation from this mean value, although we will only show plots for family I. In each case we find the critical amplitude with a precision about $\delta \phi\approx10^{-12}$.

\begin{figure}
\includegraphics[width=0.6\textwidth]{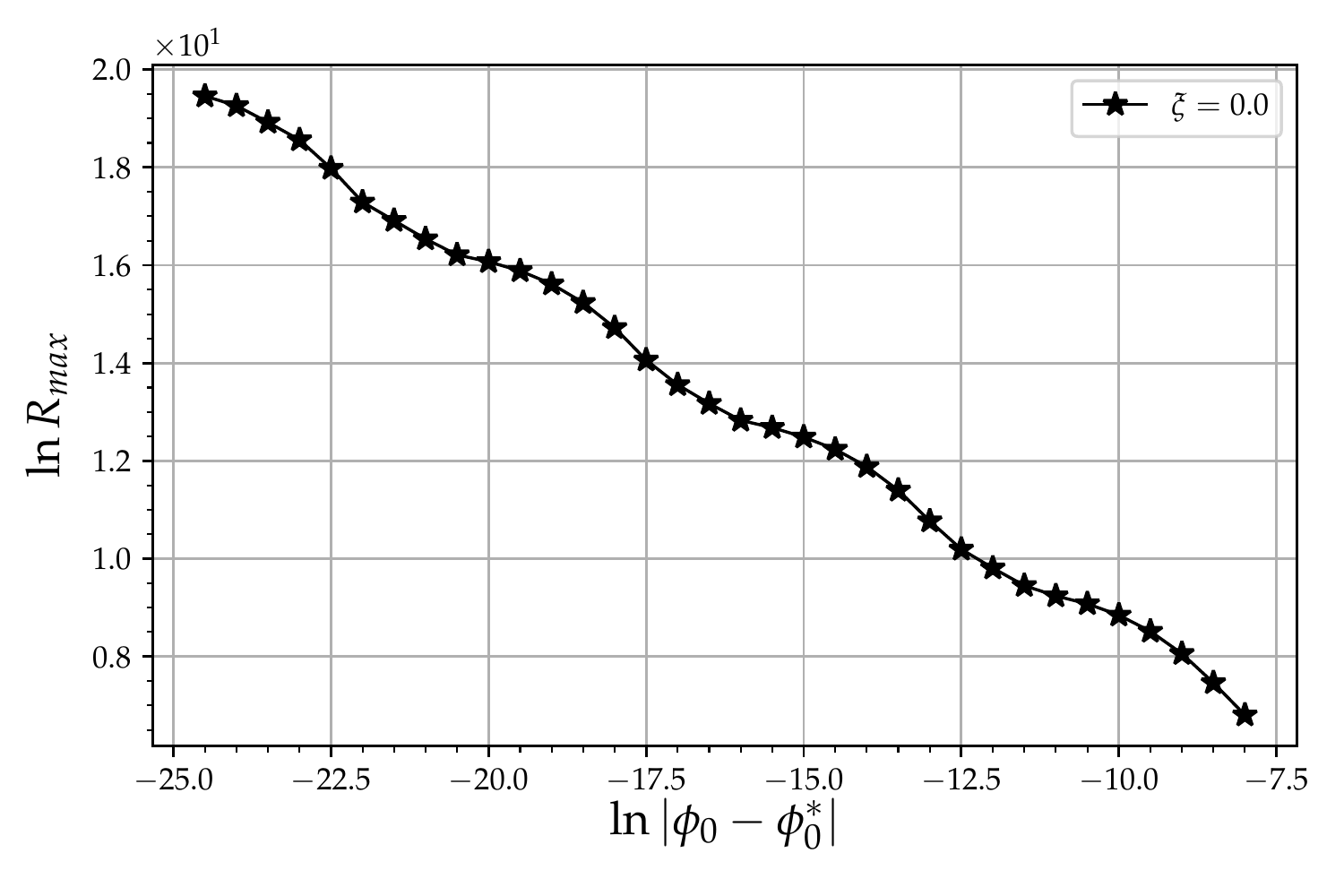}
\caption{Scaling of the maximum value at the origin of the 4D-Ricci scalar for a minimally coupled scalar field ($\xi=0$) field.  The plot corresponds to the subcritical case, and the dots are equally spaced along the $\ln|\phi_0-\phi_0^*|$ axis.}
\label{fig:lnR_xi0}
\end{figure}

\begin{figure}
\includegraphics[width=0.6\textwidth]{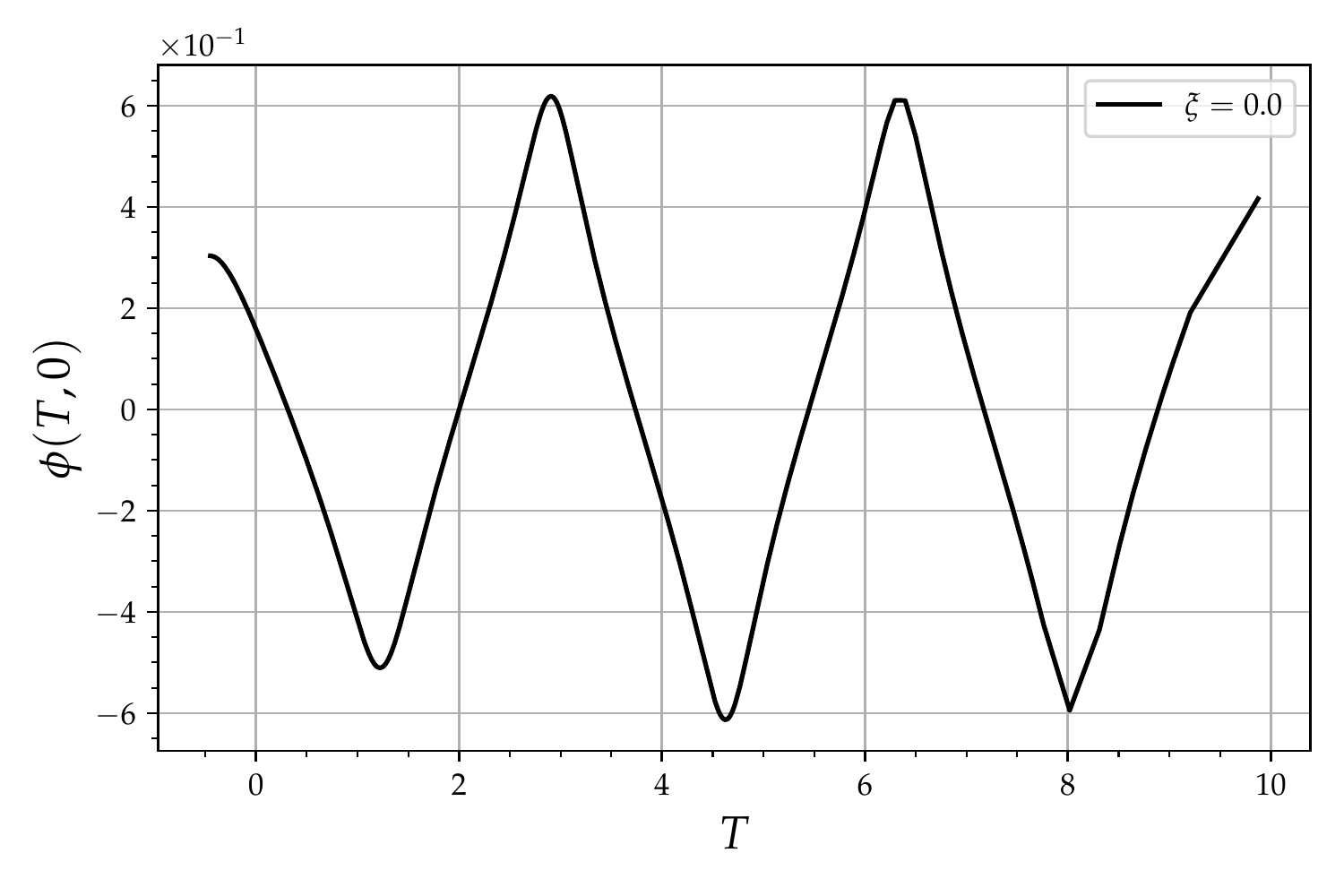}
\caption{Central value of the minimally coupled scalar field versus logarithmic time $T$, clearly showing the periodic behavior of the scalar field.}
\label{fig:echoing_xi0}
\end{figure}

In order to have a basis for comparison, we first analyze the case of a massless scalar field coupled minimally to gravity corresponding to $\xi=0$, using the 1+log slicing condition. Figure~\ref{fig:lnR_xi0} shows the maximum value of the 4D-Ricci scalar at the origin for a subcritical evolution in this case. Fitting the function~\eqref{eq:Ricci_scaling} allows us to find the critical exponents $\gamma\approx0.374\pm0.001$ and $\Delta\approx3.45\pm0.005$, which are in excellent agreement with those reported in~\cite{PhysRevD.92.084037, PhysRevD.55.695}, and for which a semi-analytical calculation gives $\gamma=0.374\pm0.001$ and $\Delta=3.4453\pm0.0005$. Figure~\ref{fig:echoing_xi0} shows the central value of the scalar field for a simulation with an initial amplitude $\phi_0=0.303350064438822$, which we are taking as the critical solution, versus the logarithmic time $T$ defined in~\eqref{eq:log_time}.  For this case we can also use the second method for calculating the echoing exponent using equations~\eqref{eq:acc_t} and~\eqref{eq:delta_echo}, obtaining $\Delta=3.42\pm 0.003$, again in good agreement with previous results. 

\vspace{5mm}

\begin{figure}[h]
\includegraphics[width=0.6\textwidth]{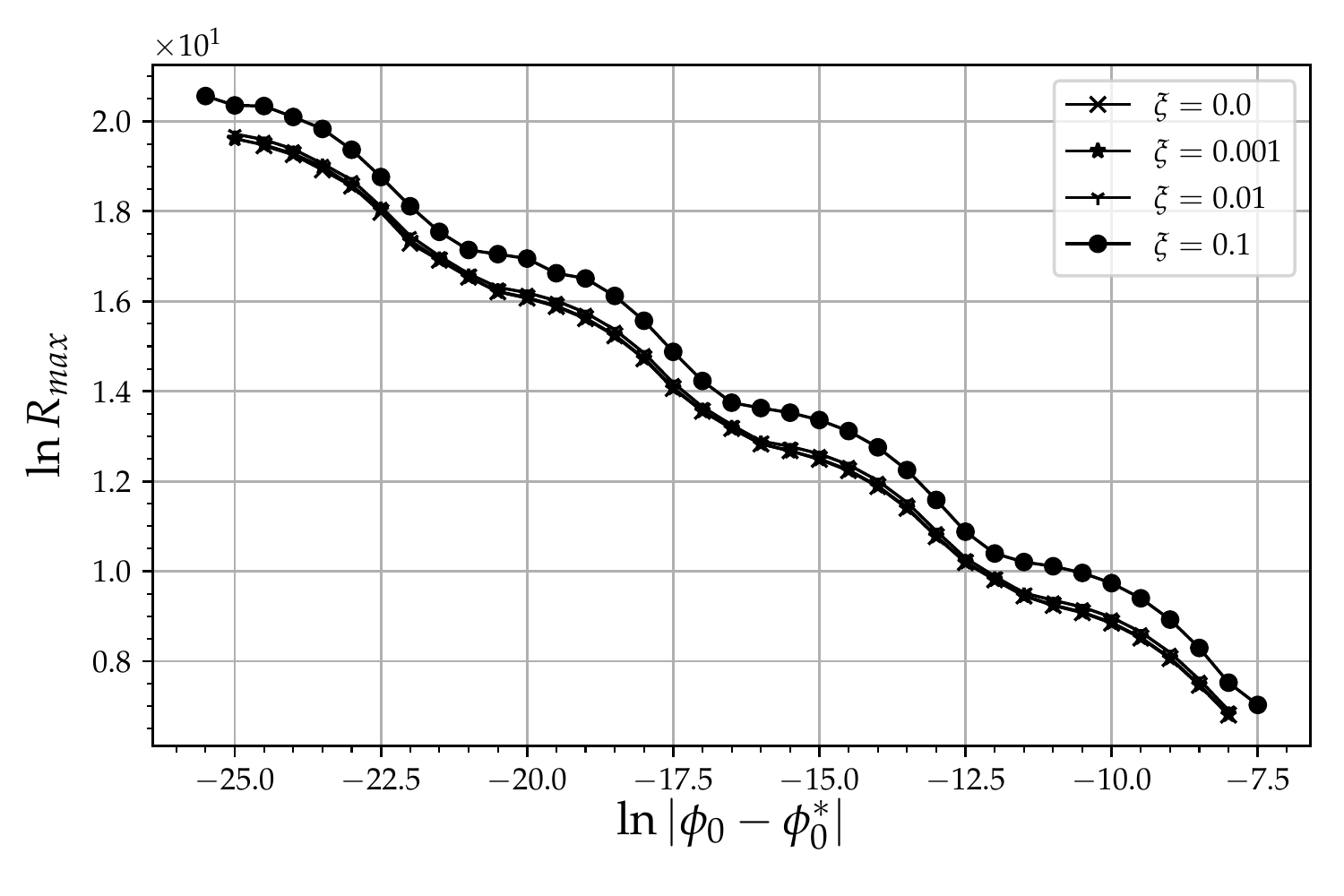}
\caption{Same as Figure~\ref{fig:lnR_xi0}, but for different values of the coupling parameter $\xi=0,10^{-3},10^{-2},10^{-1}$. All curves shows the same slope and oscillation period.}
\label{fig:lnR_xi_}
\end{figure}

\begin{figure}[h]
\includegraphics[width=0.6\textwidth]{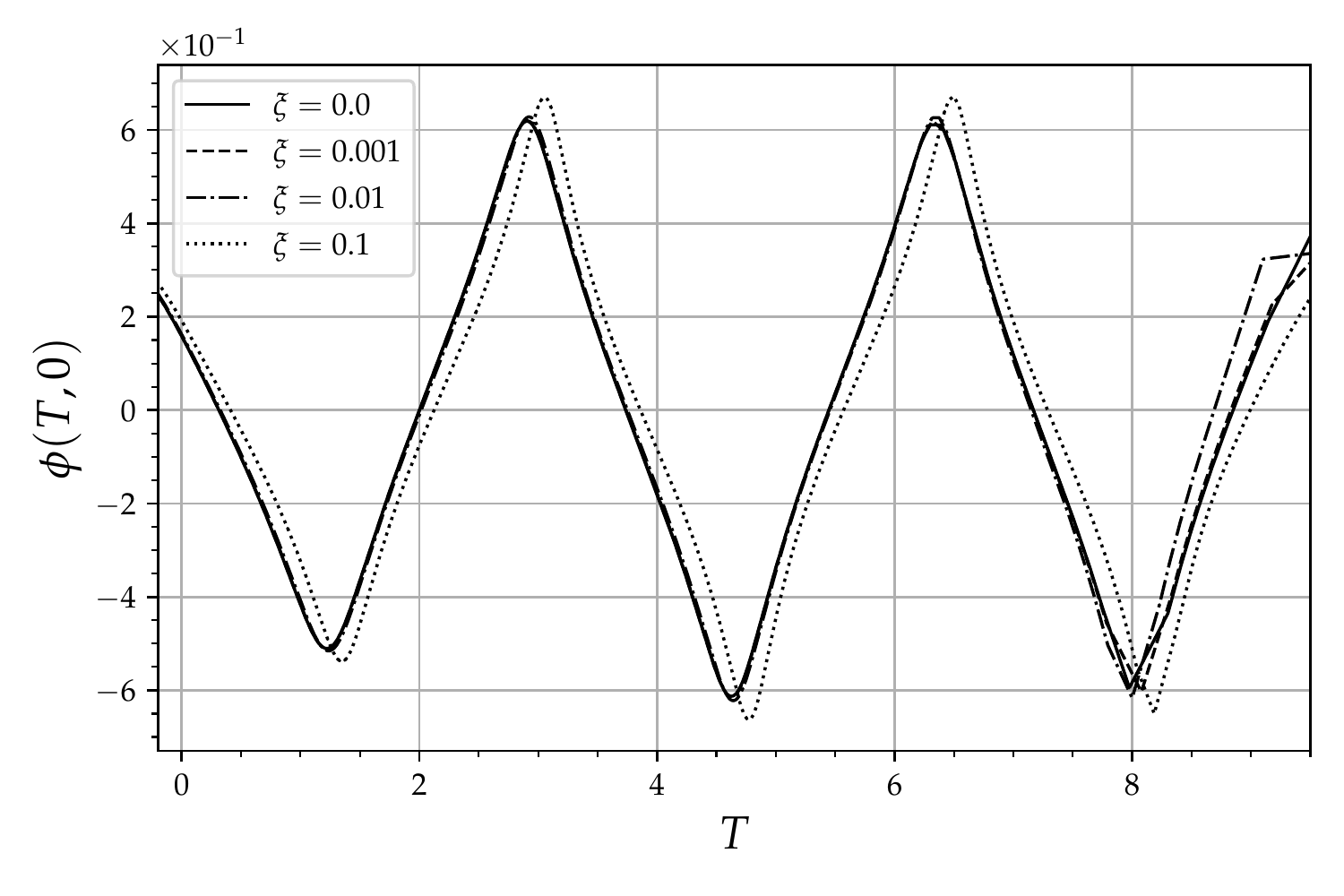}
\caption{Central value of the scalar field  plotted versus logarithmic time $T$ for $\xi=0,10^{-3},10^{-2},10^{-1}$, showing the periodic behavior of the scalar field.}
\label{fig:echoing_xi_}
\end{figure}

Next we consider the values $\xi=10^{-3},10^{-2},10^{-1}$ for the non-minimal coupling parameter in equation~\eqref{eq:nonmin_coupling}, again using the 1+log slicing condition. Results of the maximum value of the 4D-Ricci scalar at the origin for each value of $\xi$ are shown in Figure~\ref{fig:lnR_xi_}, where we also include the case $\xi=0$ for comparison. In the same way as before, we fit the function~\eqref{eq:Ricci_scaling} and obtain the critical exponents $(\gamma,\Delta)$, and we also use equations~\eqref{eq:acc_t} and~\eqref{eq:delta_echo} for a second estimate of $\Delta$. Our results are summarized in table~\ref{tab:crit_0lxil1}. Figure~\ref{fig:echoing_xi_} also shows the echoing behavior of the value of the scalar at the origin as a function of logarithmic $T$. For all these evolutions we obtain a critical exponent $\gamma \approx 0.374$, with an uncertainty of less than $0.3\%$. Also, the echoing exponent is almost exactly equal in all three cases up to a small uncertainty. We then find that there are no major differences in the critical behavior when compared to the minimally coupled case.

\begin{table}
\begin{tabular}{|c|c|c|c|}
\hline
\multicolumn{1}{|c|}{$\xi$} & \multicolumn{1}{c|}{$\gamma$} & \multicolumn{1}{c|}{$\Delta$ \eqref{eq:Ricci_scaling}} & $\Delta$ \eqref{eq:delta_echo} \\ \hline
0.001 & 0.374$\pm$0.001 & 3.441$\pm$0.001 & 3.446$\pm$0.004 \\ \hline
0.01  & 0.374$\pm$0.001 & 3.442$\pm$0.007 & 3.446$\pm$0.002 \\ \hline
0.1   & 0.372$\pm$0.004 & 3.442$\pm$0.003 & 3.445$\pm$0.005 \\ \hline
\end{tabular}
\caption{Critical exponent $\gamma$ and echoing exponent $\Delta$ (obtained by two different methods), for different values of the coupling parameter in~\eqref{eq:nonmin_coupling}. The three cases $\xi=10^{-3},10^{-2},10^{-1}$ have the same critical exponents up to a small uncertainty.}
\label{tab:crit_0lxil1}
\end{table}

\vspace{5mm}

\begin{figure}[p]
\centering
\subfloat{\includegraphics[width=0.6\textwidth]{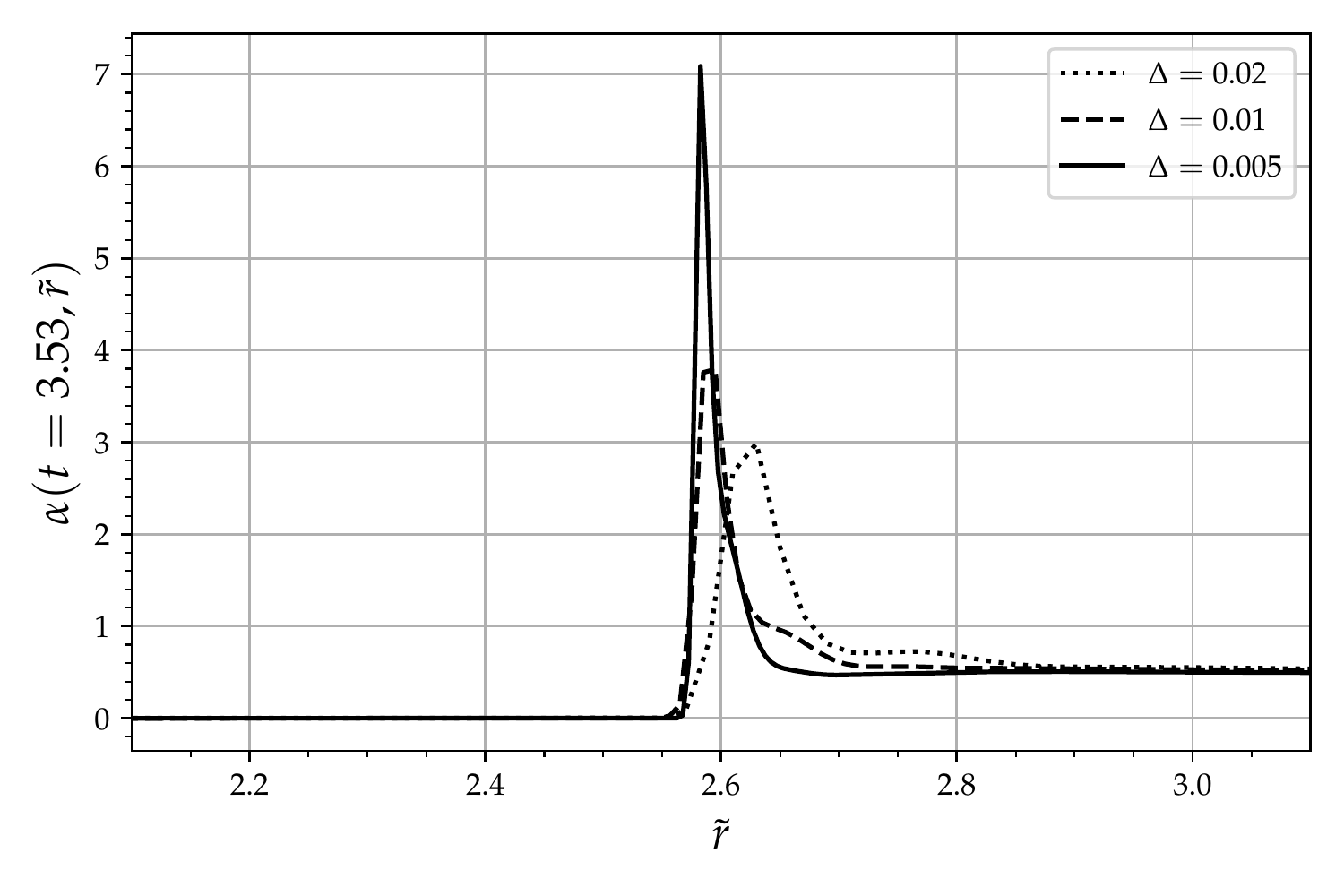}} \\
\subfloat{\includegraphics[width=0.6\textwidth]{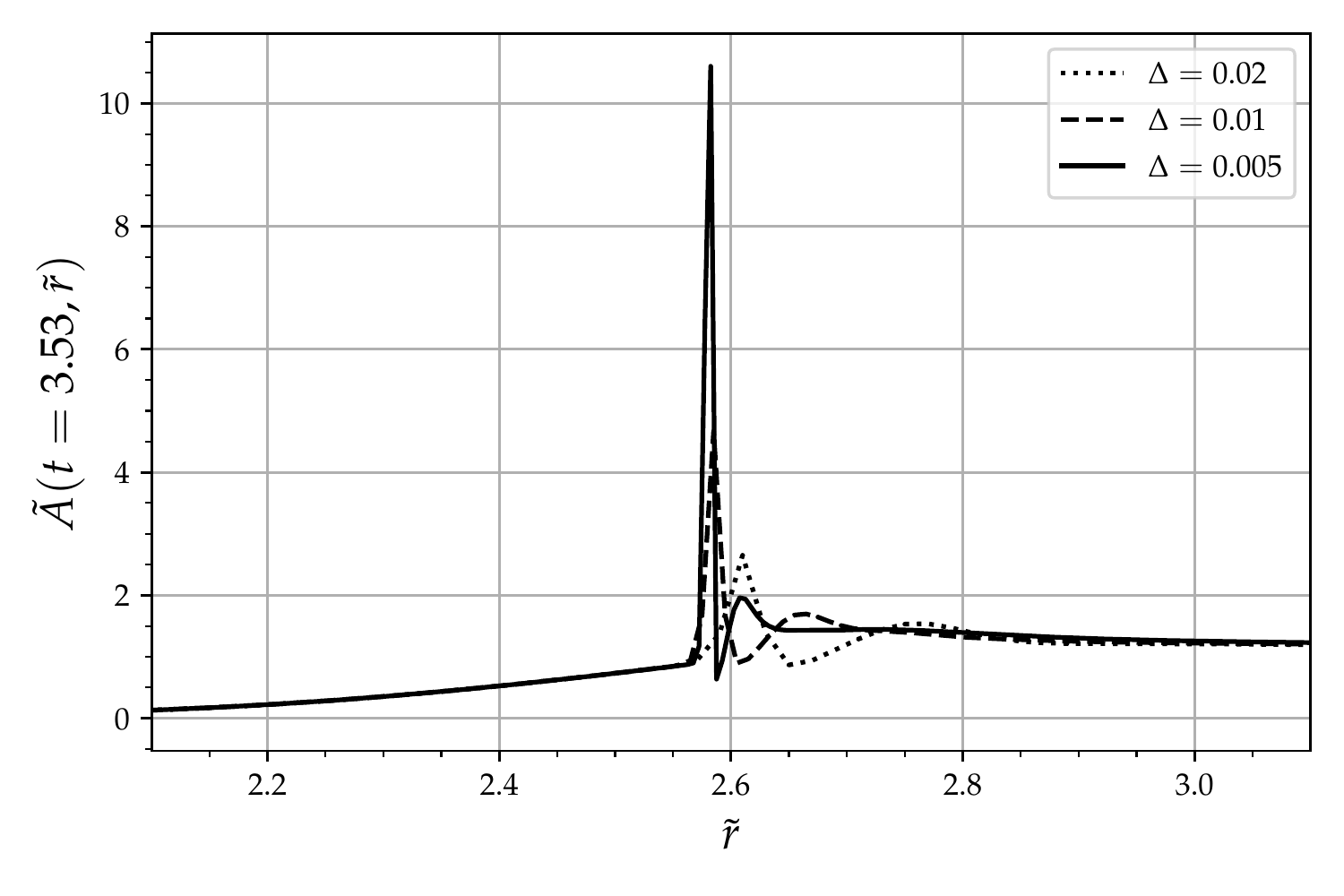}} \\
\subfloat{\includegraphics[width=0.6\textwidth]{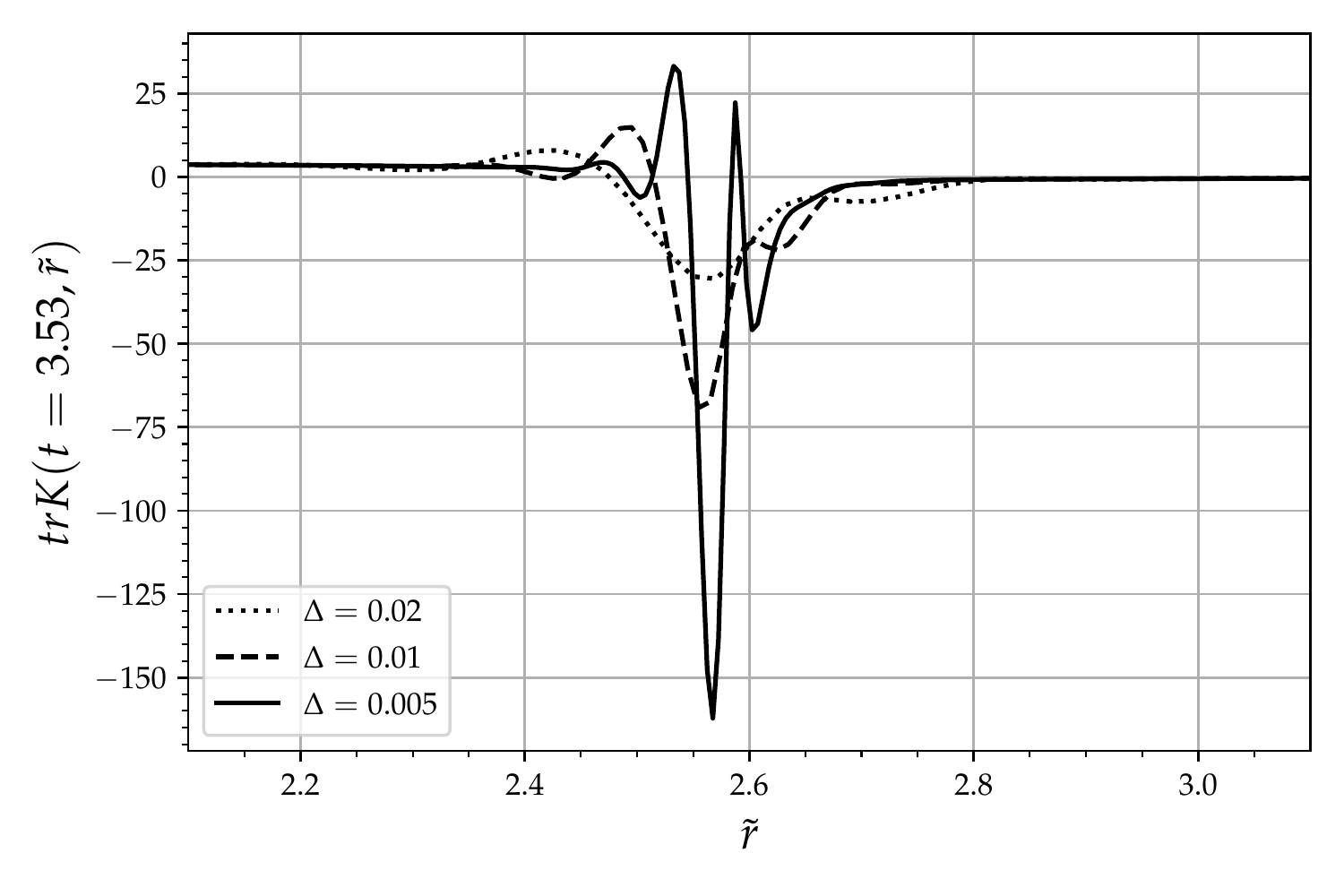}}
\caption{Snapshots of the lapse function $\alpha$, the conformal radial metric coefficient $\tilde{A}$, and the trace of the extrinsic curvature ${\rm tr}K$, at coordinate time $t\approx3.53$, for $\xi=1$ and initial data of type I with $\phi_0=0.2354$, using a 1+log slicing condition and three different resolutions. We can see that all three functions develop large gradients close to $\tilde{r}\approx2.55$, that in fact become worse with increased resolution.  These gradients cause the simulations to fail shortly after this time.}
\label{fig:peaks}
\end{figure}

Values of the coupling parameter such that $\xi \geq 1$ require a different treatment than the previous cases. For the particular case with $\xi=1$ using a 1+log slicing condition we find that, while bracketing the critical amplitude using the bisection method, once we reach a precision in the amplitude close to $\delta\phi\approx10^{-3}$ the numerical code crashes and our simulations fail. For example, when using the initial data family I (equation~\eqref{eq:gaussian}) with an initial amplitude of $\phi_0=0.2354$, we find that the lapse $\alpha$, the conformal metric coefficient $\tilde{A}$, and the trace of extrinsic curvature ${\rm tr}K$ all develop very large gradients at $\tilde{r} \approx 2.55$ that cause the code to crash at a coordinate time $t \approx 3.53$. We have found that these large gradients in fact becomes worse as we increase our numerical resolution.  This is shown in Figure~\ref{fig:peaks}, were we plot results from three different resolutions $\Delta\tilde{r}=0.02,0.01,0.005$. This behavior is quite similar to the ``gauge shocks'' described by one of the authors in~\cite{PhysRevD.55.5981,Alcubierre:2002iq}, as well as the problems reported by Hilditch {\em et al.} in~\cite{PhysRevD.88.103009} while evolving near-critical Brill wave spacetimes \cite{PhysRev.131.471,PhysRevD.16.1609}.

\begin{figure}
\centering
\subfloat{\includegraphics[width=0.6\textwidth]{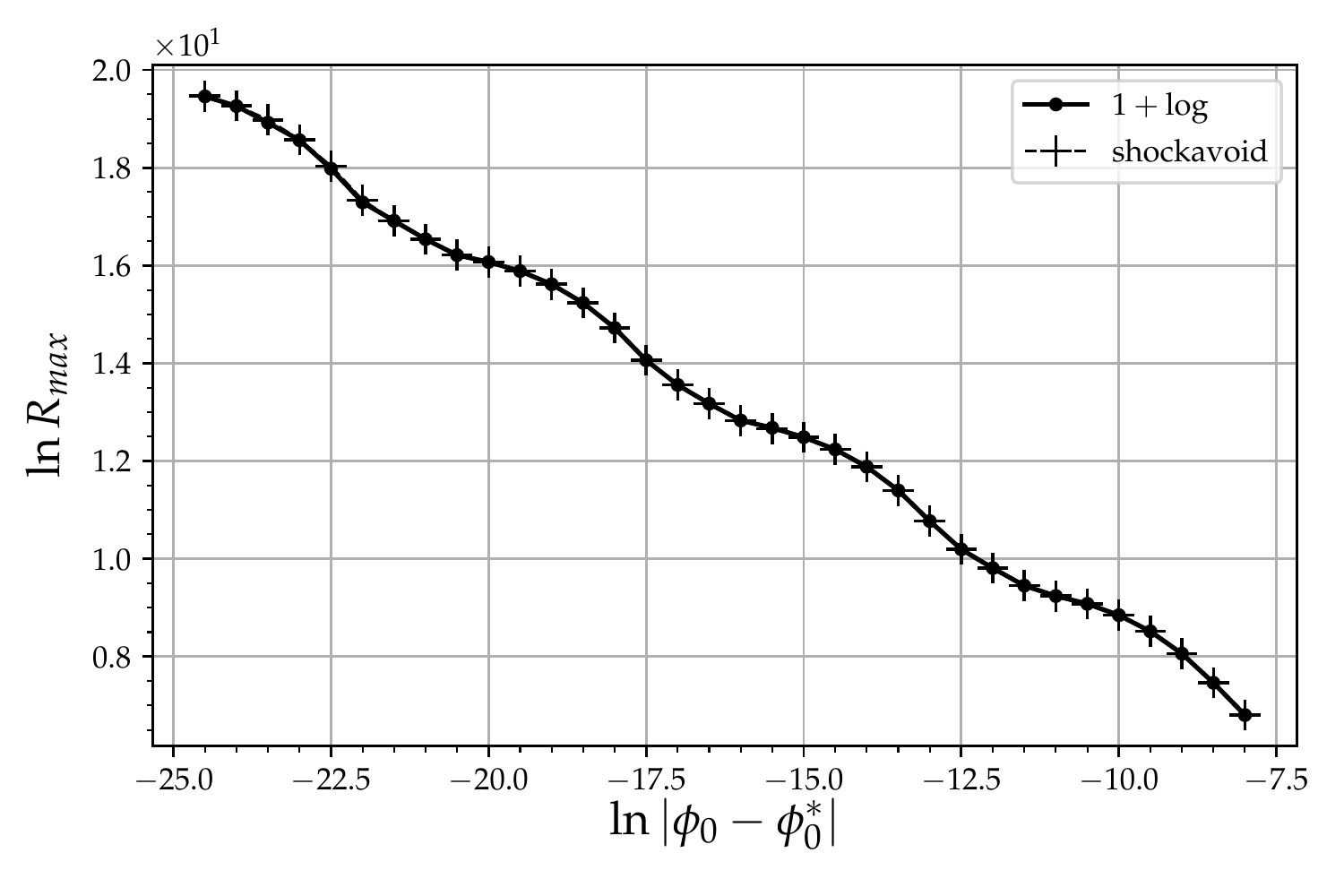}} \\
\subfloat{\includegraphics[width=0.6\textwidth]{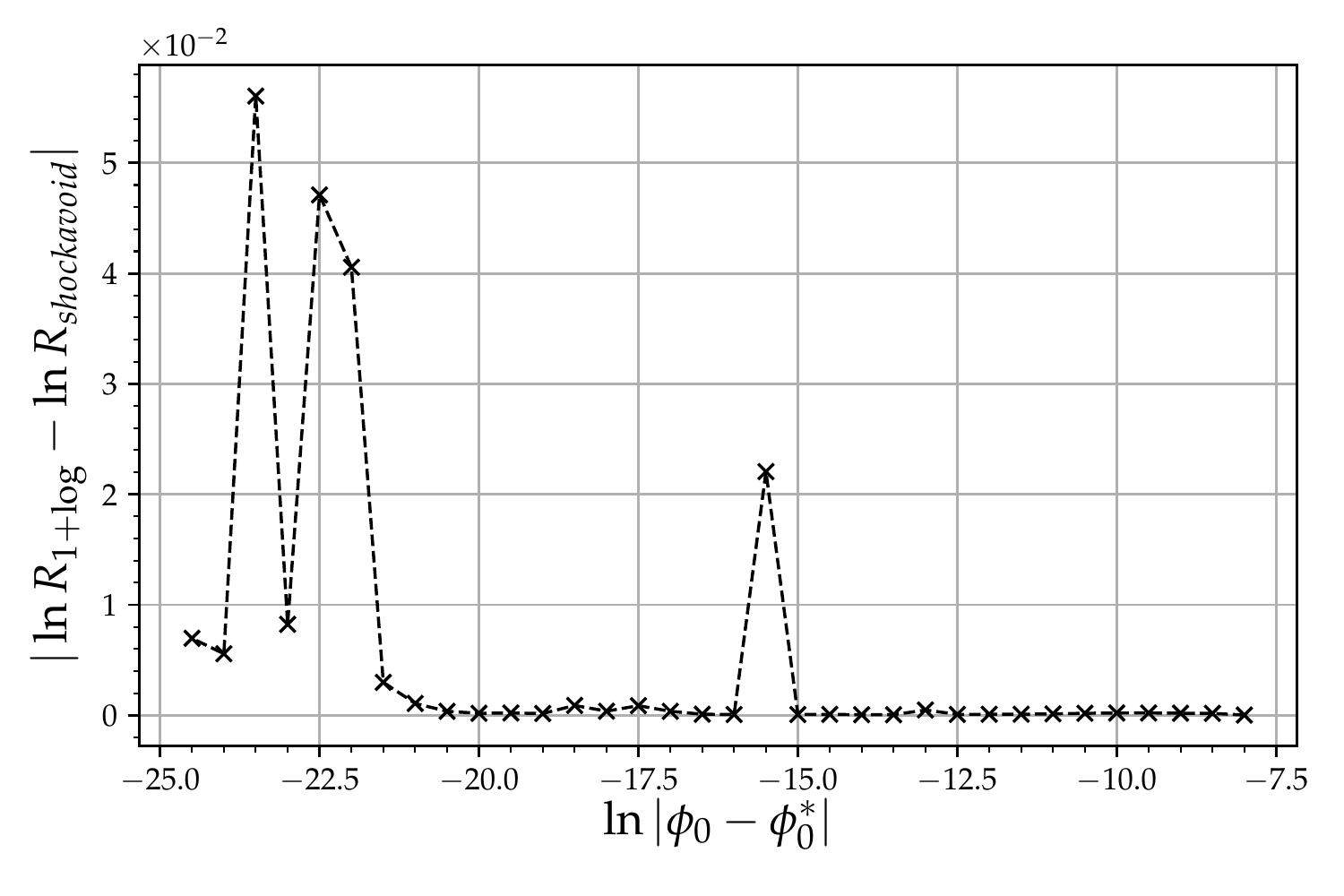}}
\caption{Top panel: 4D-Ricci scaling for the subcritical case with coupling parameter $\xi=0$, using both the $1+\log$ and shock avoiding slicing conditions. Bottom panel: Absolute value of the difference between both plots on a logarithmic scale.}
\label{fig:slicings}
\end{figure}

We have in fact traced the problem with these large gradients to the choice of the $1+\log$ slicing condition, and have therefore changed our gauge choice to the ``shock-avoiding'' lapse $f_{BM}(\alpha)=1+1/\alpha^2$ described in equation~\eqref{eq:shockavoid} (with $\kappa=1$). In order to compare both slicing conditions we also evolved the case $\xi=0$ using the shock-avoiding slicing condition, obtaining the critical exponents $\gamma=0.374\pm0.003$ and $\Delta=3.44\pm0.005$. Figure~\ref{fig:slicings} shows the critical behavior using both lapse conditions for the initial data family I.  The top panel shows a comparison of the scaling of the maximum central value of the 4D-Ricci scalar for both slicings, while the lower panel shows the absolute difference between them. We can see that for this case both slicing conditions
result in very similar evolutions.

\begin{figure}[h]
\includegraphics[width=0.6\textwidth]{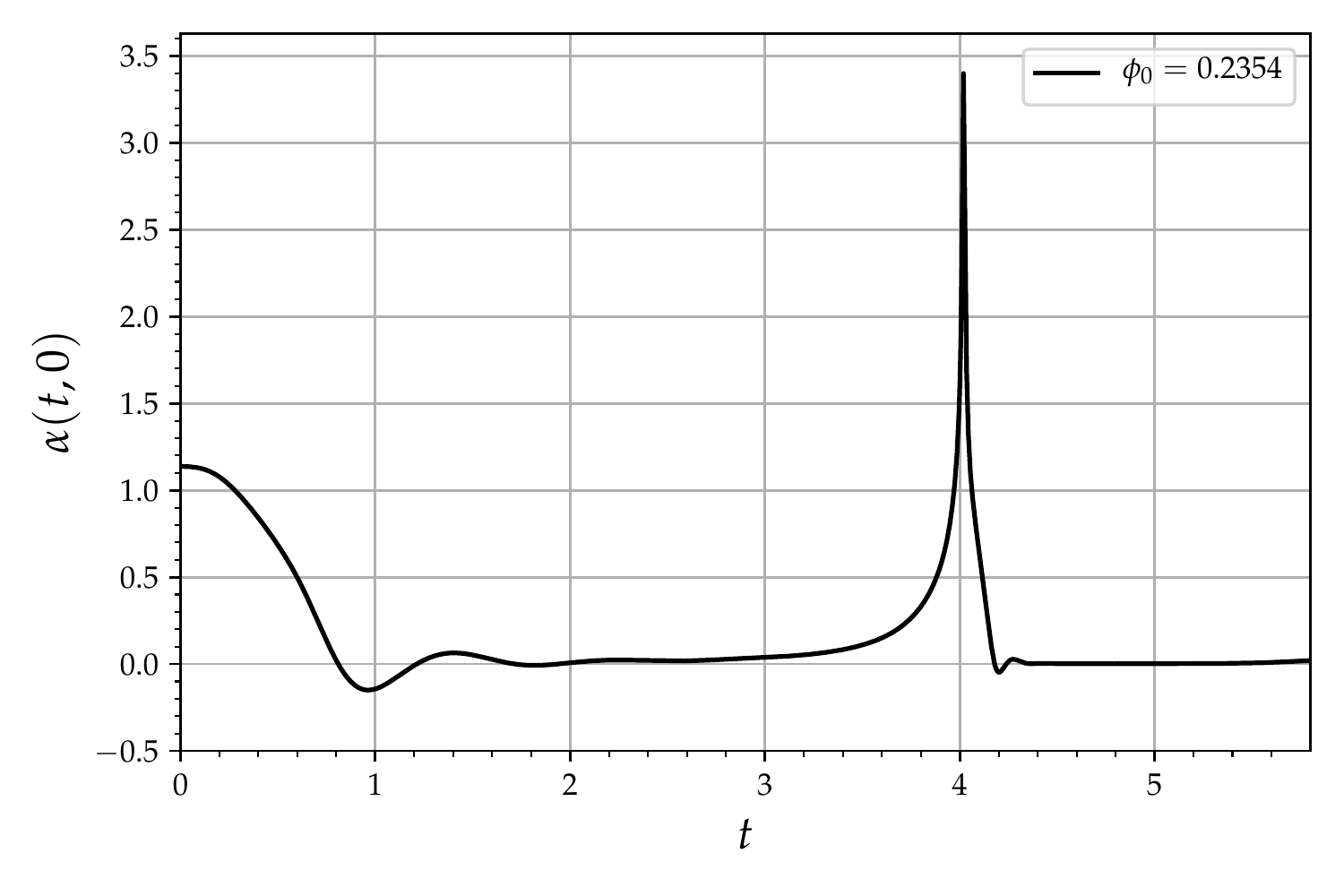}
\caption{Lapse function $\alpha$ evaluated at the origin using the shock-avoiding slicing condition for the case with $\xi=1$ and $\phi_0=0.2354$. Using this condition the lapse becomes negative at coordinate time $t\approx0.8$, but it quickly returns to positive values. The negative values of the lapse do not seem to cause any problems with the simulation.}
\label{fig:alpha_origin_shock}
\end{figure}

Changing to the shock-avoiding slicing condition now also allows us to evolve the case with $\xi=1$ and $\phi_0=0.2354$ that we mentioned above until a black hole is formed at $t\approx5.5$, thus showing that this in fact corresponds to a supercritical case. As was already pointed out in~\cite{Alcubierre:2002iq}, one possible problem with the shock-avoiding slicing condition is that lapse is now allowed to take negative values.  We can see that this is indeed the case in Figure~\ref{fig:alpha_origin_shock}, where we plot the central value of the lapse as a function of coordinate time for this same simulation.  We have found that having the lapse sometimes become negative in the central regions does not in fact seem to cause any problems.  Quite the opposite, the negative values of the lapse helps to avoid the large gradients that caused the simulations to crash with the $1+\log$ slicing condition. The negative lapse would seem to make the slices back away from a possible coordinate singularity, and later start moving forward again.

\begin{figure}
\includegraphics[width=0.6\textwidth]{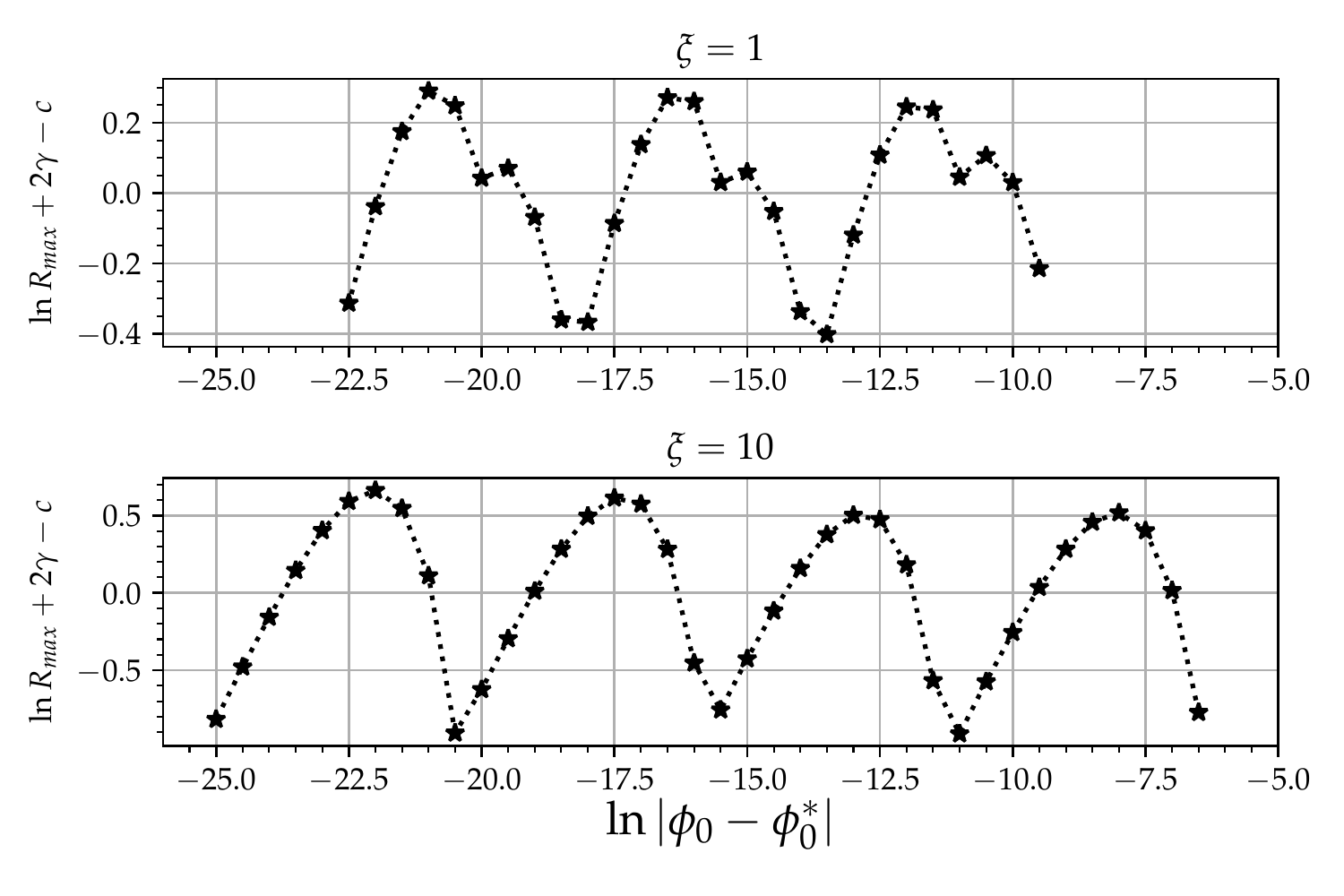}
\caption{Same data as in Figure~\ref{fig:scaling_xi=1,10} after subtracting a linear fit.  The top panel shows the case for $\xi=1$, while the bottom panel shows the case for $\xi=10$.}
\label{fig:lnR-fit}
\end{figure}

\begin{figure}[h]
\includegraphics[width=0.6\textwidth]{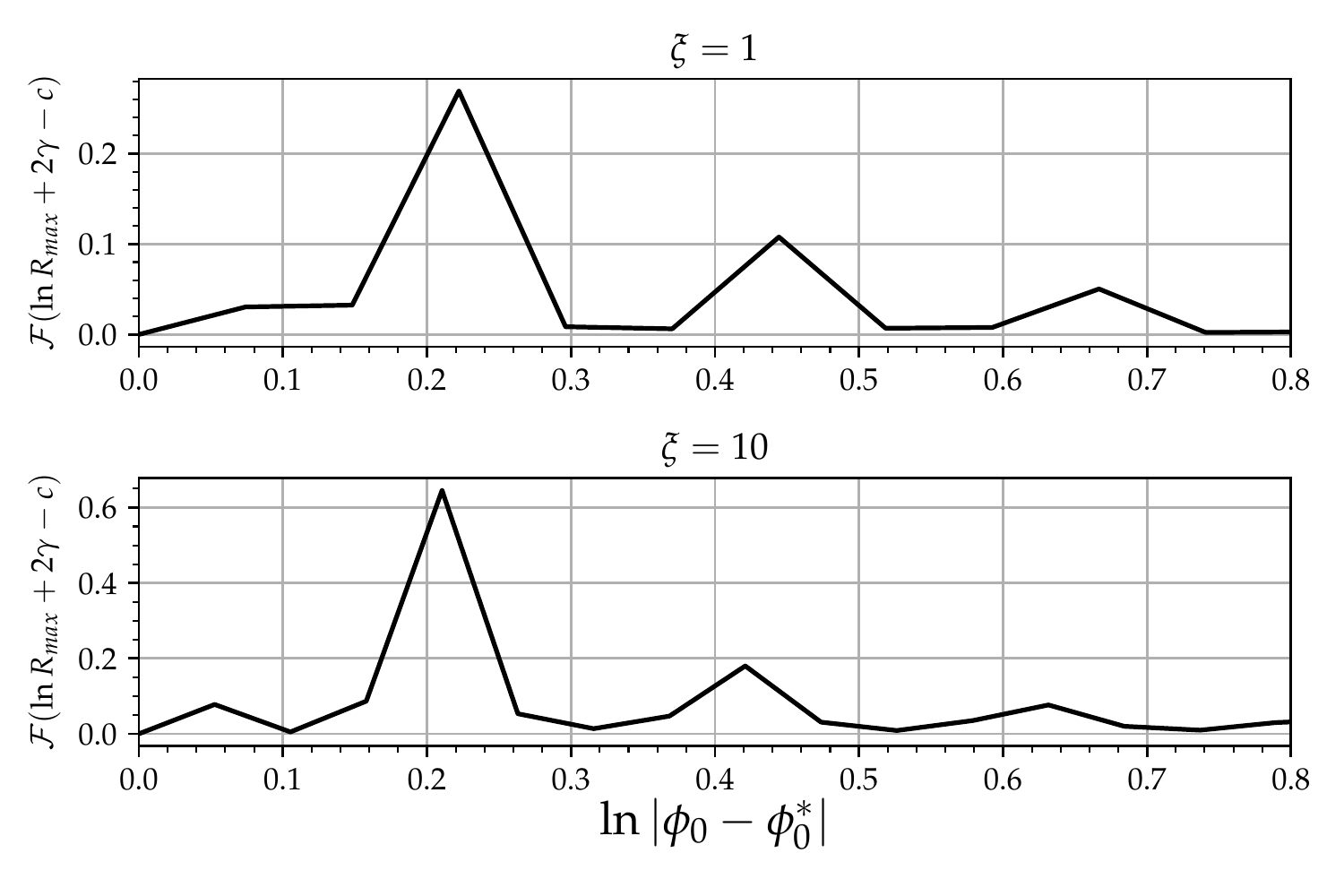}
\caption{Fourier transforms of the data shown in Figure~\ref{fig:scaling_xi=1,10}. One can clearly see a fundamental frequency plus at least its first two harmonics.}
\label{fig:frequencies}
\end{figure}

Using the shock avoiding slicing condition we have been able to perform simulations
with much larger values of the coupling parameter $\xi$, and have accurately determined the critical exponents. The critical behavior for the cases with $\xi=1,10$ is shown in Figure~\ref{fig:scaling_xi=1,10}, which plots the scaling of the maximum value of the 4D-Ricci scalar for these cases. Even by eye one can see that the plot now shows at least two different superposed oscillation frequencies.  This can be seen more clearly after subtracting a linear fit from the numerical data, as shown in Figure~\ref{fig:lnR-fit}. This observation is further confirmed by applying a fast Fourier transform (FFT) to the data after subtracting a linear fit of the form $2\gamma\ln|\phi_0-\phi_0^*|+c$. Results of this FFT can be seen in Figure~\ref{fig:frequencies}, which clearly reveals the presence of a fundamental frequency $\omega$ and at least the first two harmonics.

\begin{figure}
\includegraphics[width=0.6\textwidth]{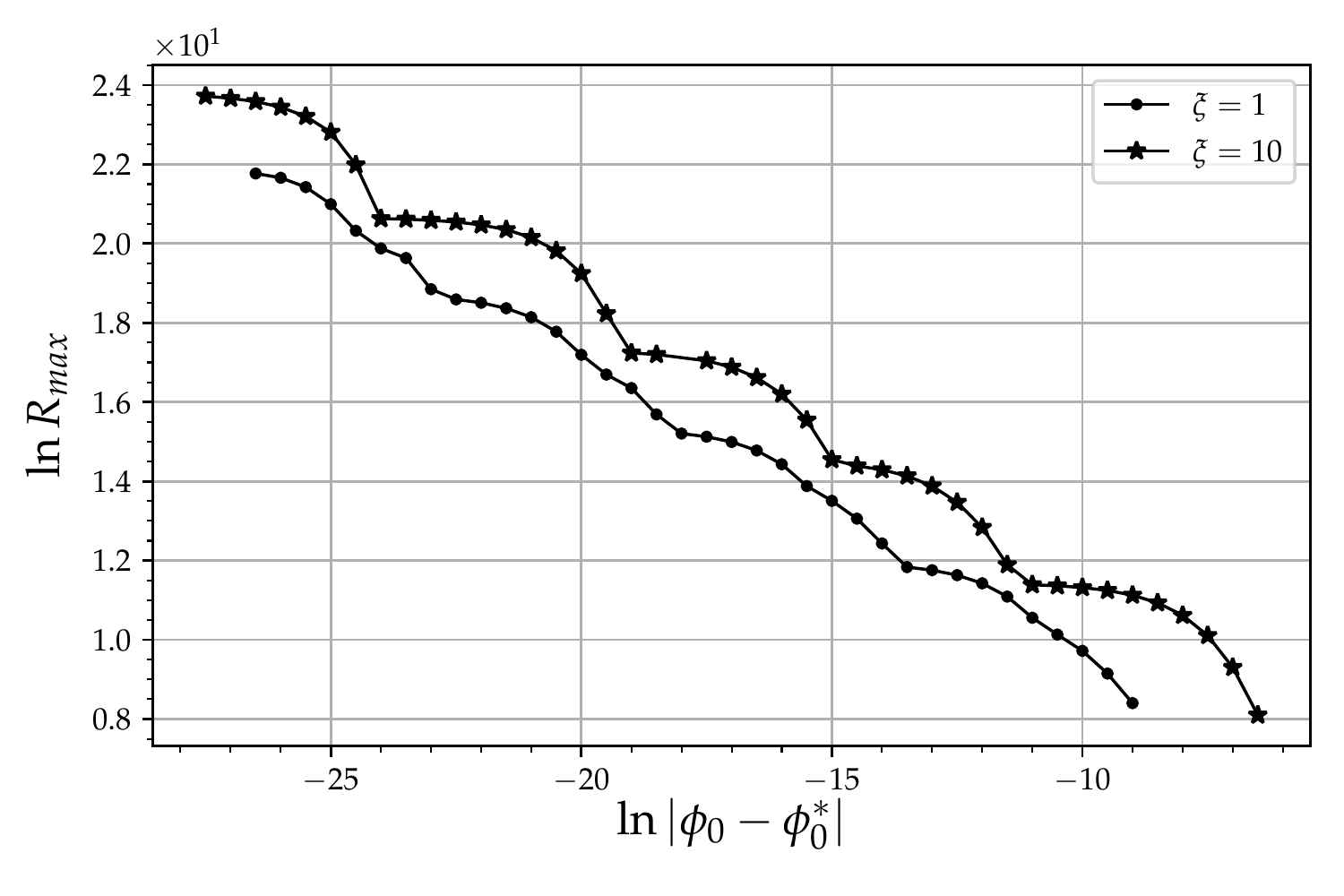}
\caption{4D-Ricci scaling for the subcritical case with coupling parameters $\xi=1,10$.}
\label{fig:scaling_xi=1,10}
\end{figure}

In fact, the difference of the 4D-Ricci scaling from the linear fit for the case with $\xi=10$ resembles a smooth sawtooth. From Figure~\ref{fig:frequencies} we can observe that the highest peaks in the Fourier transform are at frequencies $\omega$ and $2\omega$. Accordingly, instead of just fitting a function of the form~\eqref{eq:Ricci_scaling}, we add a second term with frequency $2\omega$:  
\begin{equation}\label{eq:Ricci_scaling2}
\ln R_{max} = C - 2\gamma\ln|\phi_0^*-\phi_0|+a_0\sin(\omega\ln|\phi_0^*-\phi_0|+\varphi_0)+ a_1\sin(2\omega\ln|\phi_0^*-\phi_0|+\varphi_1) \,.
\end{equation}

%Because of our limited computational resources, at this point we are not quite sure if %this extra term is due to poor numerical resolution or if is a property of the NMC %scalar field. Nevertheless, the extra term is also clearly periodic in %$\ln|\phi_0-\phi_0^*|$.

\begin{table}
\begin{tabular}{|c|c|c|c|}
\hline
\multicolumn{1}{|c|}{$\xi$} & \multicolumn{1}{c|}{$\gamma$} & \multicolumn{1}{c|}{$\Delta$ \eqref{eq:Ricci_scaling2} } & $\Delta$ \eqref{eq:delta_echo} \\ \hline
1 & 0.368$\pm$0.001 & 3.386$\pm$0.017 & 3.450$\pm$0.080 \\ \hline
10  & 0.365$\pm$0.006 & 3.109$\pm$0.007 & 2.981$\pm$0.193 \\ \hline
\end{tabular}
\caption{Critical exponent $\gamma$ and echoing exponent $\Delta$ (obtained by two different methods), for values of the coupling \eqref{eq:nonmin_coupling} $\xi=1,10$.}
\label{tab:crit_1leqxi}
\end{table}

Table~\ref{tab:crit_1leqxi} shows our results for the cases $\xi=1,10$. The uncertainty in the critical exponent $\gamma$ in each case is less than $\%2$, although the echoing exponent $\Delta$ could not be so accurately determined having an uncertainty of about $\%6$. Figure~\ref{fig:echoing_xi=1,10} also shows the periodic behavior of $\phi$ in logarithmic time $T$ for these two cases.

\begin{figure}[h]
\includegraphics[width=0.6\textwidth]{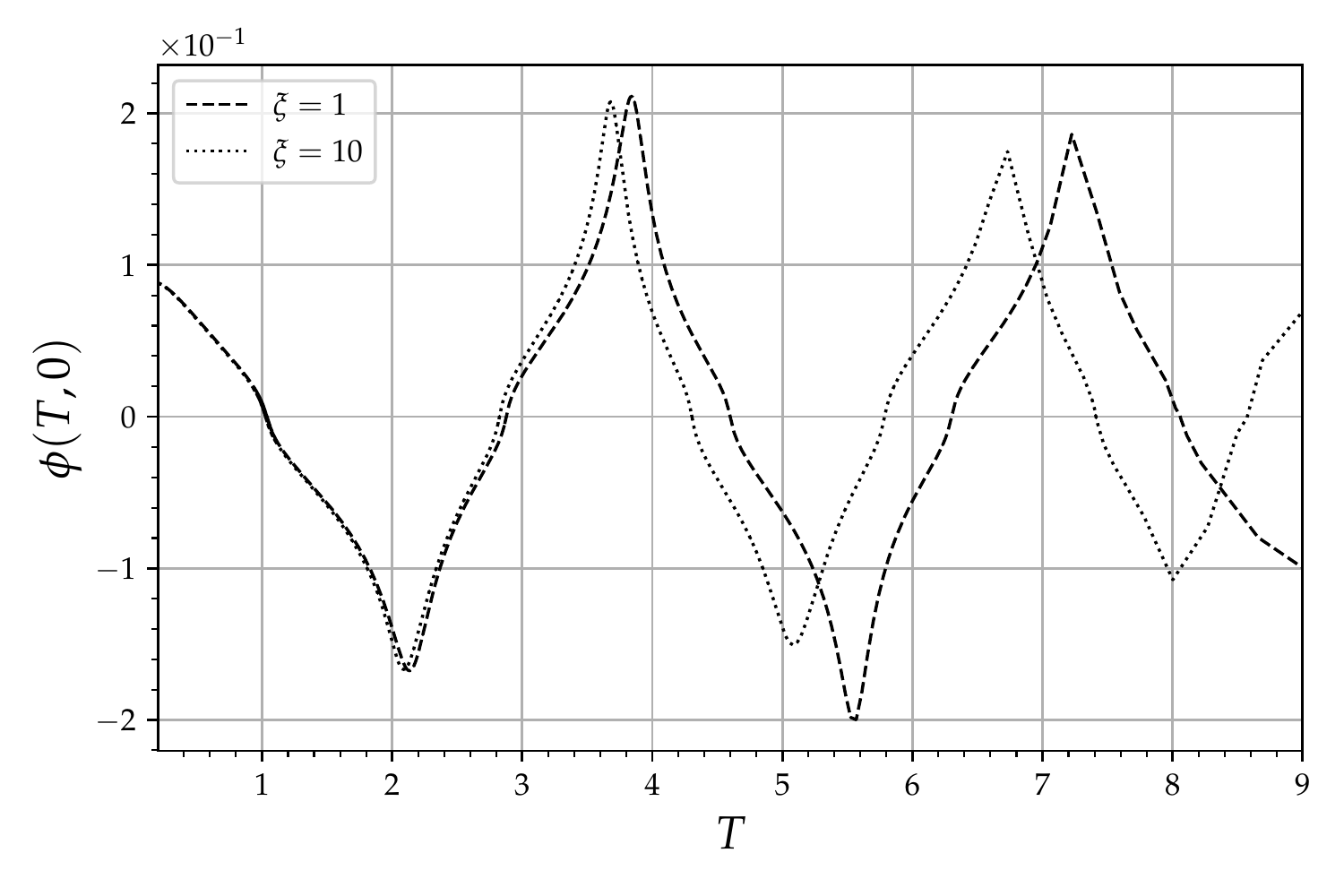}
\caption{Central value of the scalar field  plotted versus logarithmic time $T$ for $\xi=1,10$, showing the periodic behavior of the scalar field.}
\label{fig:echoing_xi=1,10}
\end{figure}

Considering only the cases showed in tables \ref{tab:crit_0lxil1} and \ref{tab:crit_1leqxi}, one can observe a decrease in the critical exponent values, even though the strong coupling cases $\xi=1,10$ require a better study. Also, for the cases with a stronger coupling it is clear that the second harmonic of the fundamental frequency cannot be neglected in the 4D-Ricci scaling.

%%%%%%%%%%%%%%%%%%%%%%%%%%%%%%%%%
%%%  SUMMARY AND CONCLUSIONS  %%%
%%%%%%%%%%%%%%%%%%%%%%%%%%%%%%%%%

\section{Conclusion}
\label{conclusion}

We performed numerical simulations for the critical collapse of a non-minimally coupled massless scalar field with a quadratic coupling function, using a BSSN code adapted to spherical symmetry. Although the original structure of the code uses a uniform radial grid, the extremely high resolution needed for evolving the data in the threshold of black hole formation required us to employ a coordinate transformation in the radial direction. With this modification we gain a factor of about $10^3$ times more resolution near the origin when compared to the original grid. In all cases the value of the critical amplitude is found to an accuracy of approximately $\delta\phi\approx10^{-12}$.

We find that for small values of the coupling parameter $\xi<1$, the results obtained for both the critical exponent $\gamma$ and echoing exponent $\Delta$ show no significant difference from the case of a massless minimally coupled scalar field. On the other hand, for values of the coupling parameter such that $\xi\geq1$ we have found that, when using the $1+\log$ slicing condition, large gradients develop in the lapse, the conformal coefficient metric $\tilde{A}$, and the trace of extrinsic curvature ${\rm tr} K$, that cause the numerical code to crash and severely limit 
the accuracy with which we can find the critical amplitude to only about $\delta\phi\approx10^{-3}$. Further analysis shows that these large gradients are even sharper at higher numerical resolutions, resembling results obtained in the study by D. Hilditch {\em et al.} on Brill waves~\cite{PhysRevD.88.103009}. For this reason, in the case of large coupling parameters we have switched to using the so-called shock-avoiding slicing condition introduced in~\cite{PhysRevD.55.5981,Alcubierre:2002iq} in order to to avoid a particular class of gauge pathologies known as ``gauge shocks".

Using this new slicing condition we have been able to follow much further the simulations with strong coupling parameters and have been able to determine the critical amplitude with high accuracy.  This has allowed us also to find both the critical exponent and echoing exponent, showing a decrease in their values as the coupling parameter $\xi$ increases. Furthermore, we have also found that for strong couplings the periodic function which appears in the scaling of the 4D-Ricci scalar could not be simply approximated by a single sine function. After performing a Fourier transform we observe that this scaling also has important contributions from the second harmonic of the fundamental frequency, and maybe even from the third. 

In summary, our results show that for small coupling parameters the simulations are very similar to the case of a minimally coupled scalar field, while for the case of large coupling parameter the evolution is significantly more complex, leading to stronger dynamics that require the use of improved gauge conditions, and resulting in a modification of the critical exponents as well as a richer periodic structure in the echoes of the scalar field.

%%%%%%%%%%%%%%%%%%%%%%%%%%%
%%%   ACKNOWLEDGMENTS   %%%
%%%%%%%%%%%%%%%%%%%%%%%%%%%

\acknowledgments

This work was partially supported by CONACyT Network Projects No. 376127  and No. 304001. EJ was also supported by a CONACyT National Graduate Grant.

%%%%%%%%%%%%%%%%%%%%%%
%%%   REFERENCES   %%%
%%%%%%%%%%%%%%%%%%%%%%

%\bibliographystyle{bibtex/prsty}
\bibliography{nonmincrit}

%%%%%%%%%%%%%%%
%%%   END   %%%
%%%%%%%%%%%%%%%

\end{document}